\documentclass[letterpaper, twocolumn, 
english,superscriptaddress,unsortedaddress,secnumarabic,nobibnotes,final,prl,floatfix]{revtex4}
\usepackage{textcomp}
\usepackage{babel}
\usepackage{units}
\usepackage{graphicx}
\usepackage{xcolor}


\begin{document}
\title{From diamond to BC8 to simple cubic and back: kinetic pathways to post-diamond carbon phases from metadynamics}
\author{Roman Marto\v{n}ák}
\email{martonak@fmph.uniba.sk}
\affiliation{Department of Experimental Physics, Comenius University, Mlynská Dolina F2, 842 48 Bratislava, Slovakia}
\author{Sergey Galitskiy}
\affiliation{University of South Florida, Tampa, Florida 33620, USA}
\author{Azat Tipeev}
\affiliation{University of South Florida, Tampa, Florida 33620, USA}
\author{Joseph M. Gonzalez}
\affiliation{University of South Florida, Tampa, Florida 33620, USA}
\author{Ivan I. Oleynik}
\email{oleynik@usf.edu}
\affiliation{University of South Florida, Tampa, Florida 33620, USA}

\begin{abstract}
The experimental observation of elusive post-diamond carbon phases 
at extreme pressures remains a major challenge in high-pressure science.
Using metadynamics with coordination-number-based collective variables
and SNAP machine-learned interatomic potential, we uncover atomistic
mechanisms governing the transformation of cubic and hexagonal
diamond into post-diamond phases above 1.5 TPa. The transition initiates via homogeneous nucleation of nanoscale
liquid droplets, which rapidly crystallize into either BC8 (below 1.8 TPa) or simple
cubic phases (above 2.1 TPa), once the liquid nucleus surpasses a critical size. 
 Favorable conditions for synthesizing BC8 are identified near
1.8 TPa and 3500--5000 K. Decompression pathways from simple cubic and BC8 phases were also simulated to study possible experimental 
recovery of post-diamond carbon allotropes at ambient conditions. We also find a new metastable low-enthalpy structure with four-coordinated carbon atoms and space group \textit{P222}. Our  insights provide a theoretical foundation for experimental discovery of ultra-dense carbon phases under extreme conditions.

\end{abstract}
\maketitle
\begin{figure}[htpb]
\centering
\includegraphics[width=0.85\columnwidth]{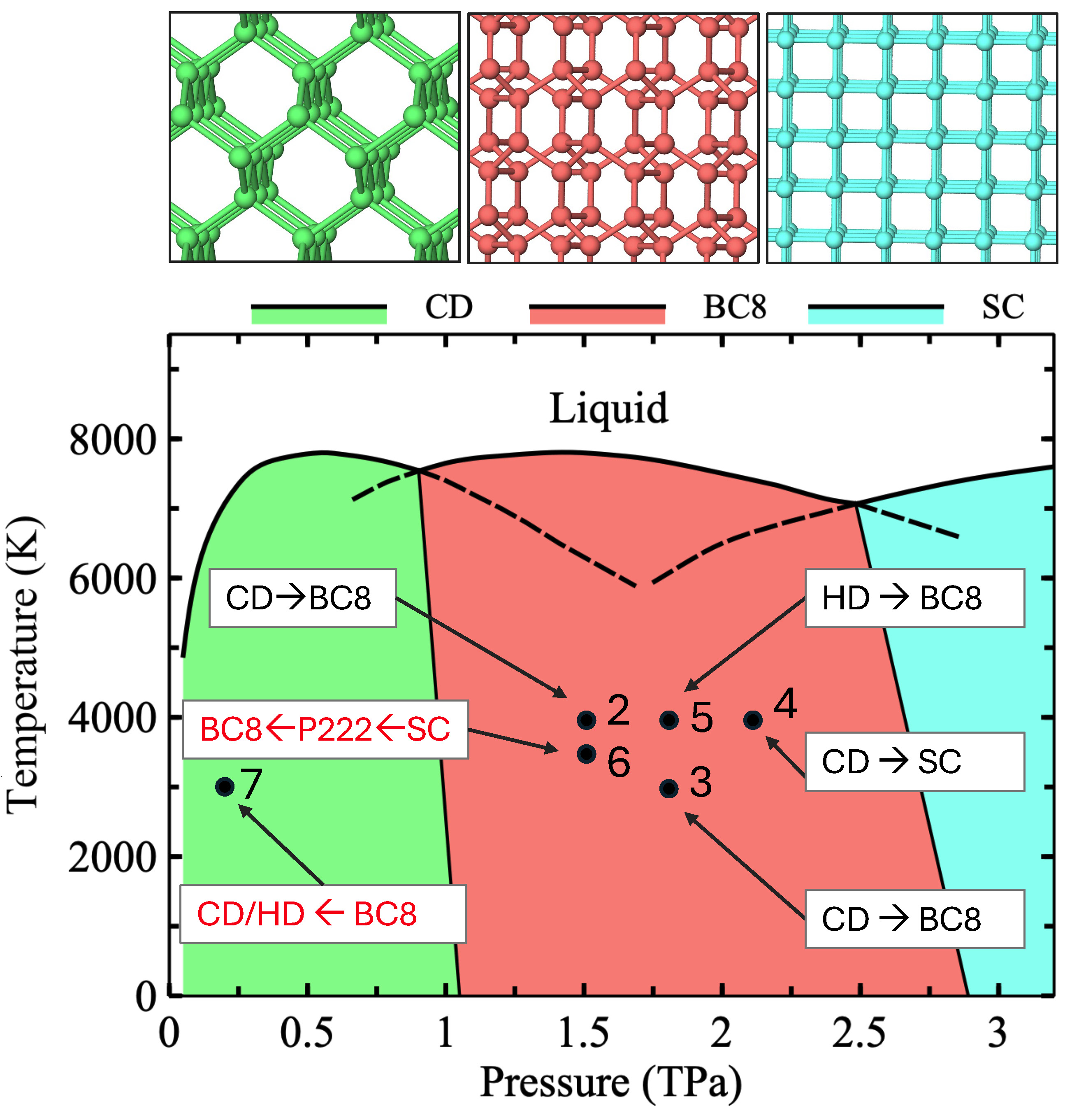}
\caption{Phase diagram of carbon showing equilibrium phase boundaries and predicted transition pathways. 
Metastable extensions of the solid-liquid phase boundaries are shown as dashed lines. Numbers next to black circles refer to the structural transitions shown in Figures 2-7. The compression and decompression pathways are labeled by black and red colors, respectively.}
\label{fig:kinetic_phase_diag}
\end{figure}

The diamond phase of carbon is renowned not only for its exceptional
physical properties, such as extreme hardness, outstanding thermal
conductivity, and optical transparency \citep{Field1992,Field2012}, but also for its remarkable
thermodynamic stability across an exceptionally wide range of pressures
and temperatures \citep{Lazicki2021,Smith2014,Nguyen-Cong2024}. Post-diamond phases of carbon, first predicted by Yin and Cohen \citep{Yin_Cohen1983,Yin1984}, include the 8-atom body-centered (BC8)
structure, stable between 1 and 3 TPa, and the simple
cubic (SC) phase, stable above 3 TPa at $T=0$ (Fig.~\ref{fig:kinetic_phase_diag}) \cite{Martinez-Canales2012}. However, experimental validation of these predictions remains extremely challenging \cite{Smith2014,Lazicki2021}.

Both experiments \citep{Lazicki2021,Smith2014} and simulations \citep{Nguyen-Cong2024}
reveal the extreme metastability of cubic diamond (CD) at pressures up to 2
TPa, well beyond its thermodynamic phase boundary at temperatures
below 4000 K. The absence of a transition to the thermodynamically stable
BC8 phase above 1 TPa is attributed to the very strong covalent
bonding in diamond, which creates a large kinetic barrier between
the crystalline phases. A double shock compression pathway to BC8 phase was proposed based on molecular dynamics (MD) simulations \cite{Sun2023}. A review of studies of diamond at extreme conditions is given in Ref.~\cite{review2024}.

Very recently, a novel diamond-to-BC8 transformation mechanism 
has been uncovered in large-scale MD simulations using a quantum-accurate Spectral Neighbor Analysis
Potential (SNAP) \citep{Willman2022} and an initial nanocrystalline
diamond (NCD) sample  \cite{Nguyen-Cong2024}. It involves an appearance of a metastable liquid carbon intermediate, which allows to overcome a significant crystal-to-crystal transition barrier by melting diamond first, followed by nucleation and growth of BC8 from the
supercooled carbon liquid.  The presence of grain boundaries in the NCD sample significantly facilitated melting. This raises an important question: can such melting and subsequent BC8 nucleation be observed in single-crystal diamond samples, which lack extended defects to seed the transformation, and on much longer
time scales beyond the reach of conventional MD?

A powerful technique to study the fundamental mechanisms of solid--solid
phase transformations is metadynamics \citep{metadynamics} -- an
enhanced sampling method that efficiently explores free energy landscapes
and helps overcome barriers in MD simulations of rare events  (for a review see Ref.~\cite{Bussi2020}).
This approach is particularly useful for studying first-order structural transitions
in crystals \citep{PRL2003,Martonak2005,natmat2006,OganovBook,Hromadova2013, Plasienka2015,Badin2021}, where activated processes are inaccessible to standard MD, as the transitions occur on timescales too long to be sampled. Metadynamics adds to the interaction potential of the system a history-dependent bias potential that acts on a selected set of collective variables (CVs)  -- low-dimensional descriptors of the system's important degrees of freedom relevant to the structural transition. 
Early studies employed supercell vectors \cite{PRL2003,Martonak2005,natmat2006} or average coordination number \cite{Zipoli2004, Hu2015} as CVs while recently more complex CVs were constructed \cite{Piaggi2017,Piaggi2018,PiaggiEnthalpyPRL2017}, including neural-network-based path collective variables \cite{Rogal2019}.
In Ref.\cite{Badin2021} it was shown that the average coordination number is a simple, generic, and efficient CV that not only captures displacive mechanisms of pressure-induced structural transformations but also enables access to nucleation processes in sufficiently large systems. Very recently, coordination number based CV was also used to study complex structural transformations in coesite\citep{vrba2025}.

An early metadynamics study \cite{Sun2009} using \textit{ab initio} MD and supercell vectors as CVs \cite{natmat2006} did not observe the 
diamond-to-BC8 transformations. At 2 TPa and 4000 K, CD transformed to the SC phase rather than
to BC8. The BC8 phase was observed only upon subsequent decompression
of the SC phase to 1 TPa at 5000 K. The limitation of that study
was the very small system size (64 atoms), which significantly restricted the exploration of phase space. 

In this work, we take the next step forward by employing the recently
developed quantum-accurate SNAP model
for carbon \citep{Willman2022}, which enables metadynamics simulations
of much larger systems and longer timescales \cite{Nguyen-Cong2021,Willman2024}. By focusing on
phase transformations in compressed and heated single-crystal
diamond, our study complements direct SNAP MD simulations of phase
transformations in NCD \citep{Nguyen-Cong2024} and extends the prior simulations of Sun et al.~\citep{Sun2009}
to include homogeneous nucleation mechanisms of new carbon phases.
The goal is to determine the pressure--temperature conditions under
which cubic (CD, cF8) and closely related metastable hexagonal diamond (HD, hP4, lonsdaleite) transform into BC8. In addition, we explore potential
transformations induced by decompression of the SC and BC8 phases and
address the important question of whether BC8
can remain metastable upon decompression to lower pressures and potentially
be recoverable at ambient conditions.

The metadynamics investigation of diamond-to-post-diamond transformations
was performed using two coordination numbers as CVs -- focusing at the first and second coordination shells 
in carbon phases (for more detailed information on the CVs, see description and Fig.~S1 in Supplemental Material \cite{SM} (see also 
Refs.~\cite{LAMMPS,PLUMED,OVITO,mlsi,Perdew1996,Kresse1993,Kresse1996a,Kresse1996b,Kresse1999,MonkhorstPack,VESTA} therein)).
The MD and metadynamics simulations were performed with the LAMMPS \cite{LAMMPS} and PLUMED \cite{PLUMED} packages.
The use of the  SNAP potential \citep{Willman2022}
enables simulations of relatively large atomic systems up to
4,096 atoms across a wide range of pressures (1.5--2.1 TPa)
and temperatures (2000--5000 K). As shown in Ref.~\citep{Badin2021}, system size plays a crucial role: its increase  results in a change in the fundamental mechanism of the phase transition, from martensitic,
collective atomic rearrangements to nucleation and growth. We now present in Figs.~2-7 (prepared by the OVITO package\citep{OVITO}) typical kinetic pathways observed upon compression and decompression at specific pressure-temperature points, shown at the phase diagram in Fig.~\ref{fig:kinetic_phase_diag}. The time evolution of CVs,  fractions of the relevant phases, as well as the exploration of the space of CVs in all presented simulations are provided in  Figs.~S2-S7 of Supplemental Material \cite{SM}.
The identification of structural phases was performed using the Machine Learning Structure Identifier (MLSI) algorithm described in Ref.~\cite{mlsi} (see also Supplemental Material \cite{SM}).

\begin{figure}[htpb]
    \centering
    \includegraphics[width=0.95\columnwidth]{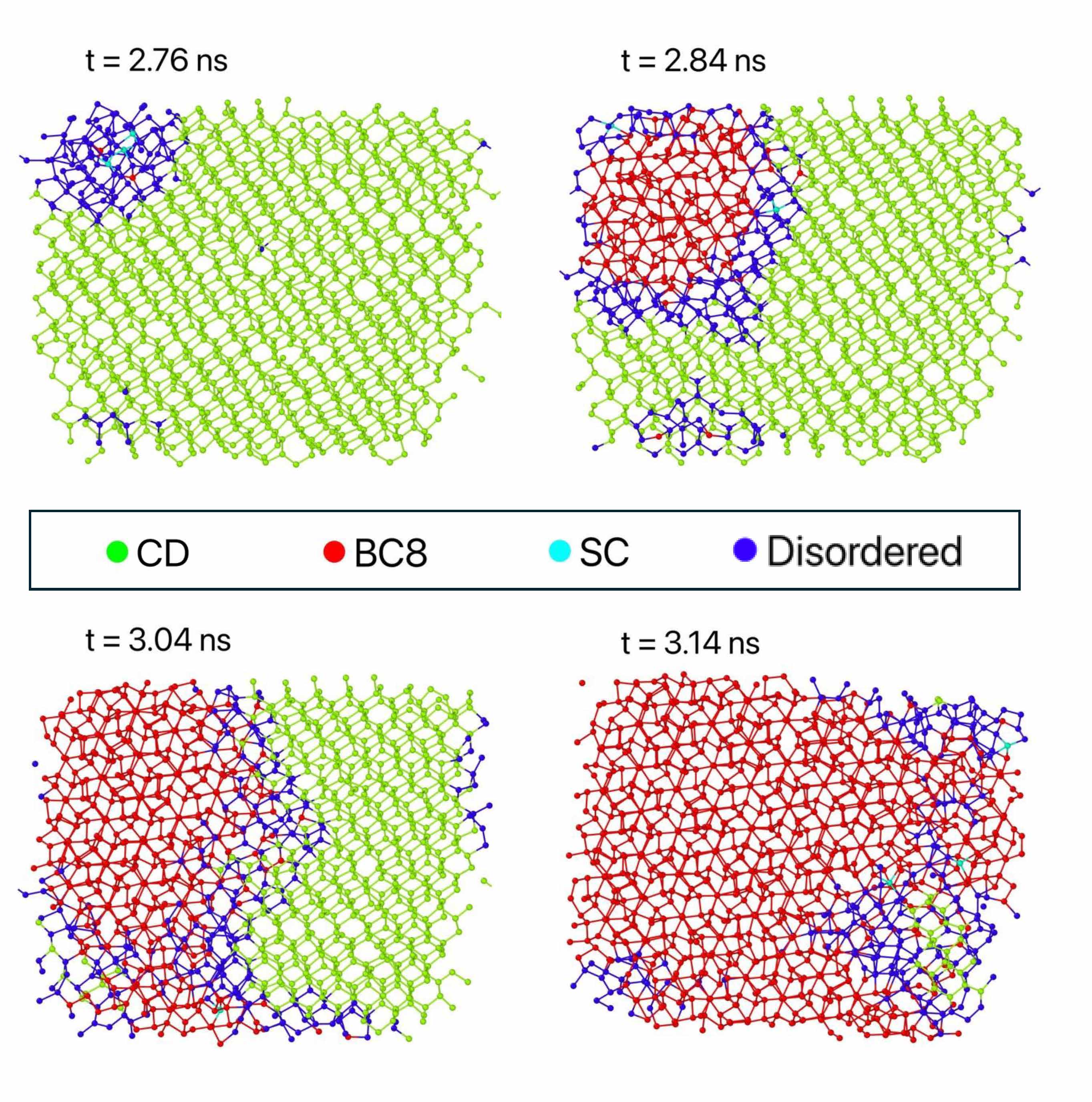}
    \caption{Transformation of CD to BC8 at 1.5 TPa and 4000 K. A disordered
    (amorphous/liquid) droplet first emerges at 2.76 ns, followed by its
    crystallization into BC8 at 2.84 ns. The BC8 phase continues to grow at 3.04 ns, leading to an almost complete transformation by
    3.14 ns.  The time evolution of the phase fractions is shown in Fig.~S2.} 
    \label{fig:cd_to_bc8_15}
\end{figure}

At 1.5 TPa and temperatures between 3500 and 5000 K, CD
transforms into BC8 via nucleation of an intermediate
disordered region. In a representative case at 4000 K (Fig.~\ref{fig:cd_to_bc8_15}, Fig.~S2),
a localized disordered zone repeatedly forms and recrystallizes back to diamond
until it exceeds a critical size of $\sim$ 350 atoms at
$\unit[2.7]{ns}$, when BC8 rapidly nucleates and grows within
the droplet. This transient disordered state acts as a precursor,
lowering the barrier for BC8 emergence. Complete conversion to a BC8
nucleus occurs by $\unit[2.9]{ns}$, followed by  formation of an interface
with the parent diamond lattice. The nucleus then expands within the
matrix, fully consuming the initial phase by $\unit[3.2]{ns}$ (Fig.~\ref{fig:cd_to_bc8_15}).
The estimated free-energy barrier of $\sim$ 500 eV (see Supplemental Material \cite{SM} for information about the barrier calculation) is substantial
but consistent with expectations for homogeneous nucleation in a defect-free and 
strongly bonded crystal. At a lower temperature of 3000 K, however,
no phase transition is observed even over 7 ns of metadynamics simulation. Although transient disorder and spontaneous crystallization into simple motifs such as SC or hexagonal occur, the system ultimately
reverts to diamond, indicating that nucleation of the liquid precursor is thermodynamically suppressed at lower temperatures.

\begin{widetext}
\begin{figure*}[htpb]
\includegraphics[width=0.75\paperwidth]{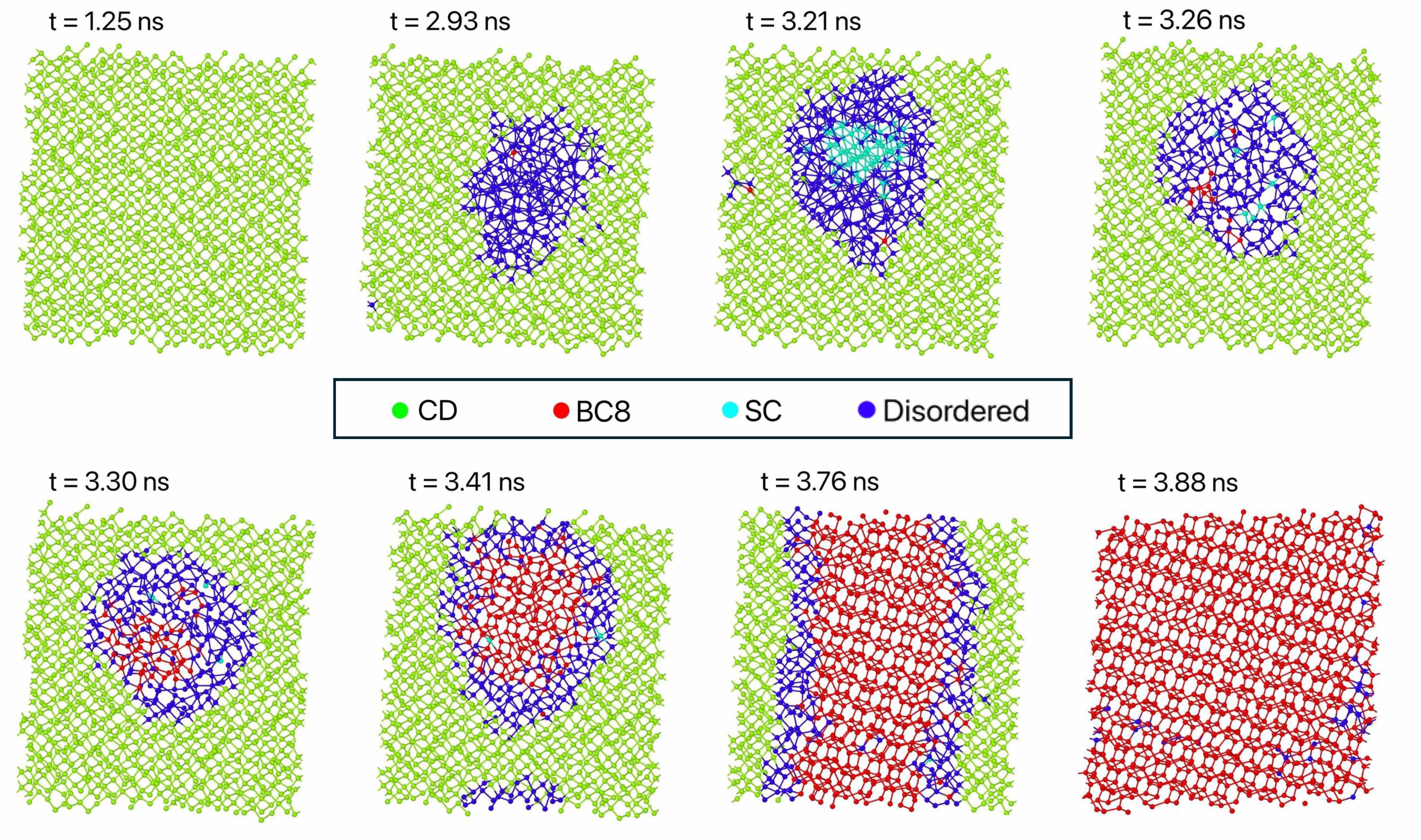}
\caption{Structural evolution of the disordered droplet during the transformation
of CD to BC8 at 1.8 TPa and 3000 K. The droplet forms at
2.93 ns, followed by transient crystallization into a SC structure
at 3.21 ns and subsequent melt at 3.26 ns. A BC8 nucleus then emerges within the disordered phase at 3.3 ns, followed by its growth and near-complete transformation to BC8 by 3.88 ns. 
The time evolution of the phase fractions is shown in Fig.~S3.}
\label{fig:cd_to_bc8_18}
\end{figure*}
\end{widetext}

At a higher pressure of 1.8 TPa and temperatures from 4000 to 5000
K, a similar transformation to BC8 is observed. However, at lower temperatures
(down to 2500 K), a metastable SC structure becomes more
prevalent. For example, at 3000 K, a disordered droplet is also formed, but it first crystallizes into an SC crystal, followed by melting and recrystallization into BC8 (Fig.~\ref{fig:cd_to_bc8_18} and Fig.~S3). Growth then proceeds rapidly, fully consuming the parent phase. The transient appearance of the SC phase can be attributed to its lower structural complexity (one atom per primitive cell vs. eight in BC8) and increased energetic accessibility under elevated pressure, making it a kinetically favorable intermediate.
At 2500 K we observe a similar sequence of events,
while at 2000 K, the diamond phase remains stable and no transformation is observed.

\begin{figure}[htpb]
\centering
\includegraphics[width=0.95\columnwidth]{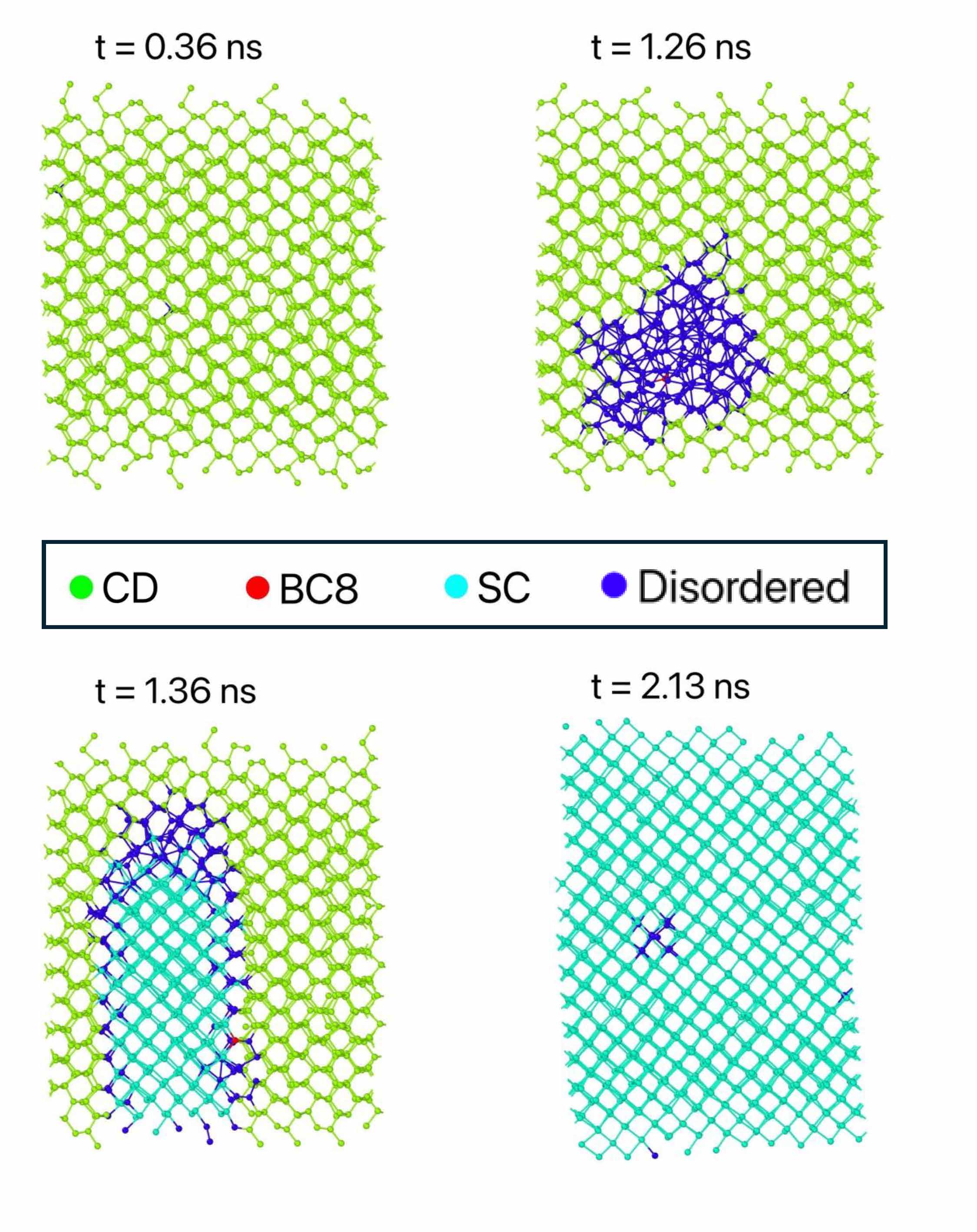}
\caption{Transformation of CD to the SC phase at 2.1 TPa
and 4000 K. A disordered droplet forms at 1.26 ns, followed by its
crystallization into the SC phase at 1.36 ns. The SC
phase grows first into a layer which persists for about 800 ps and then a complete transformation is achieved.  The time evolution of the phase fractions is shown in Fig.~S4.}
\label{fig:cd_to_sc}
\end{figure}

Upon increasing the pressure to 2.1 TPa, while sampling temperatures between 4500 and 5000 K, the BC8 phase forms through a disordered droplet. However,
at lower temperatures between 2500 and 4000 K the
transformation instead favors  the SC phase over BC8. This behavior at 4000 K is illustrated  in Fig.~\ref{fig:cd_to_sc} and Fig.~S4. Initially, a disordered region emerges and crystallizes into an SC cluster containing up to 450 atoms, which subsequently reverts
to diamond. In a second nucleation event, however, a SC nucleus
forms within a liquid-like droplet and irreversibly grows into a slab
of SC crystal, eventually consuming the entire simulation
volume. 

\begin{figure}[htpb]
\includegraphics[width=0.95\columnwidth]{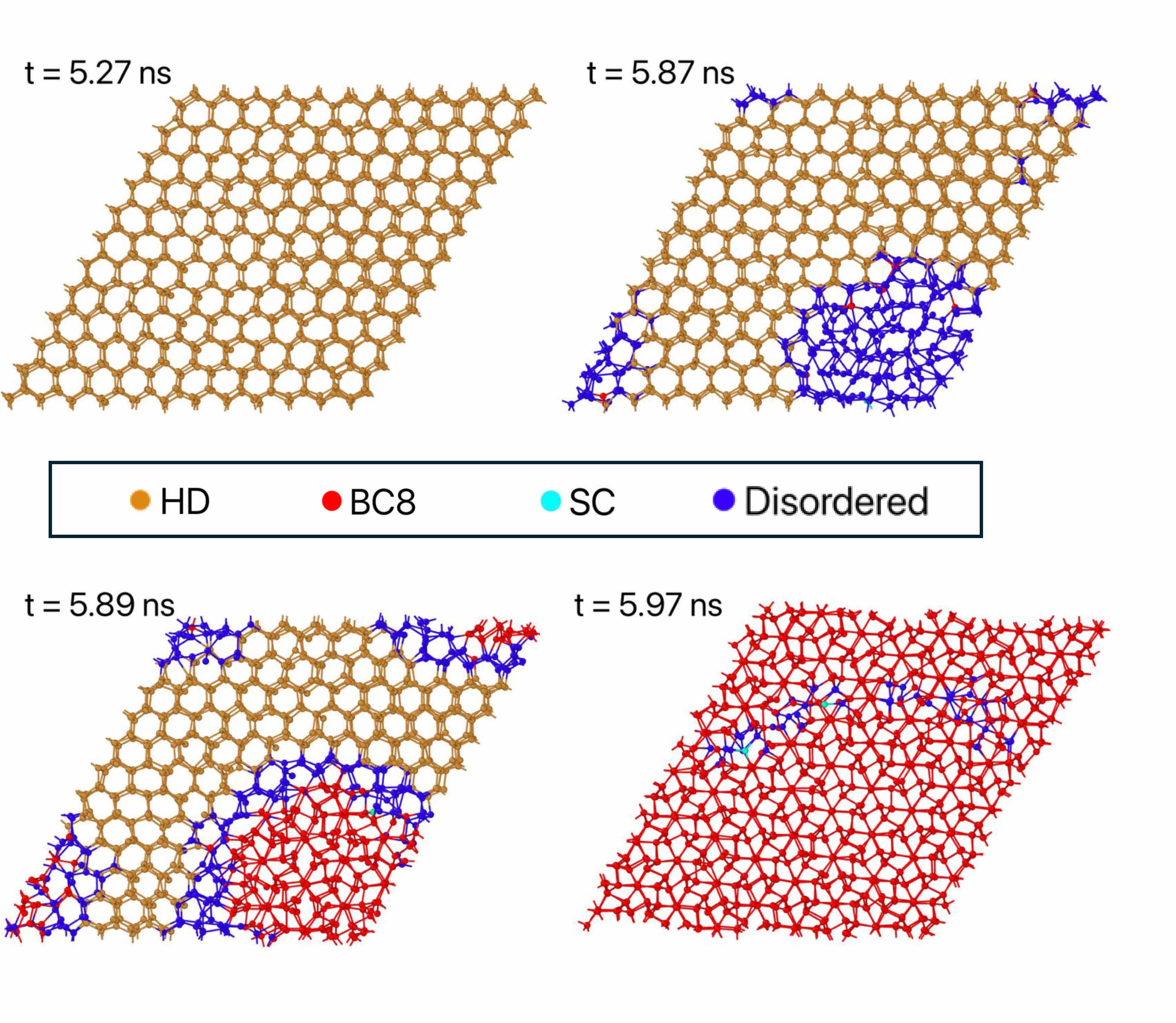}
\caption{Transformation of HD to BC8 at 1.8 TPa and 4000 K.
A disordered nucleus forms within the parent hexagonal
diamond structure at 5.87 ns, followed by its transformation into the
BC8 phase 8 ps later. Near-complete formation of the BC8 phase occurs
after 72 ps.  The time evolution of the phase fractions is shown in Fig.~S5. }
\label{fig:hd_to_bc8}
\end{figure}

We also performed a similar series of simulations starting from metastable HD, since, to our knowledge, this transition pathway has not yet been explored. At 1.8 TPa and 4000 K, we again observed the initial nucleation of a liquid-like droplet. In this case, however, crystallization into the BC8 phase occurred rapidly, within $\sim$ 8 ps. The BC8 phase then grew quickly to consume the entire hexagonal lattice within 72 ps (Fig.\ref{fig:hd_to_bc8} and Fig.~S5). In this case, the fast crystallization from the droplet preserved the orientational relationship between the parent and product crystal structures: the [111] direction of BC8 is parallel to the hexagonal axis of HD, but different orientations were also observed.
The observed fast crystallization may indicate that BC8 could be more readily synthesized experimentally from HD than from CD. At a higher pressure of 2.1 TPa, as in the case of CD, HD transformed into the SC phase.

\begin{figure}[htpb]
\includegraphics[width=0.95\columnwidth]{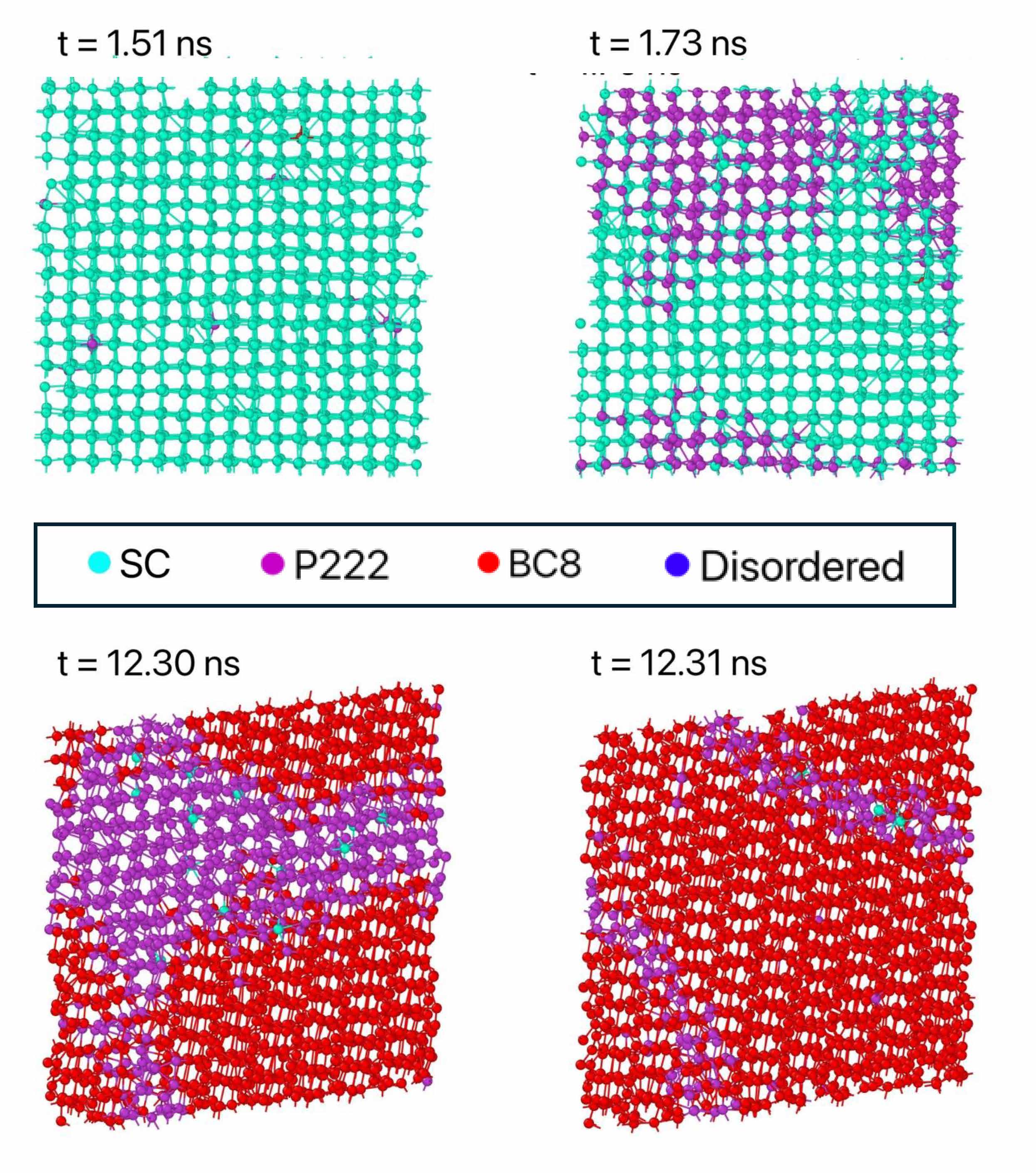}
\caption{Transformation of the SC phase to BC8 upon decompression
to 1.5 TPa and 3500 K. Onset of the transition to the intermediate
\textit{P222} phase at 1.73 ns is followed by its growth into a defective crystal, formation
of fully developed \textit{P222} phase at 8.18 ns, finally followed by a transformation
of the \textit{P222} phase into BC8 at 12.31 ns.  The time evolution of the phase fractions is shown in Fig.~S6. }
\label{fig:sc_to_bc8}
\end{figure}

Our findings therefore suggest that the pressure range in which BC8 can form
directly from diamond is relatively narrow. Under excessive overpressurization beyond 2 TPa,
the system preferentially stabilizes in the SC phase, bypassing
BC8. To further explore the diversity of structural transformations,
we examined the potential reverse transformations of SC 
phase upon its decompression to 1.5 TPa at various temperatures. The
evolution at 3500 K is shown in Fig.~\ref{fig:sc_to_bc8} and Fig.~S6. 
Surprisingly, we found  that the sample first transforms
into a novel crystalline intermediate, which subsequently converts
into BC8. Upon quenching to $T=0$, this intermediate is identified
as a metastable crystal with a complex four-coordinated structure with 16 atoms in the unit cell and \textit{P222} symmetry (see Supplemental Material \cite{SM}
for details of the structure (Fig.~S8), including enthalpy differences, and electronic density of states (Fig.~S9)). These results indicate that even if diamond is driven
to the SC phase by extreme compression, BC8 can still emerge
upon subsequent decompression, consistent with prior observations \cite{Sun2009}.

\begin{figure}[htpb]
\includegraphics[width=0.95\columnwidth]{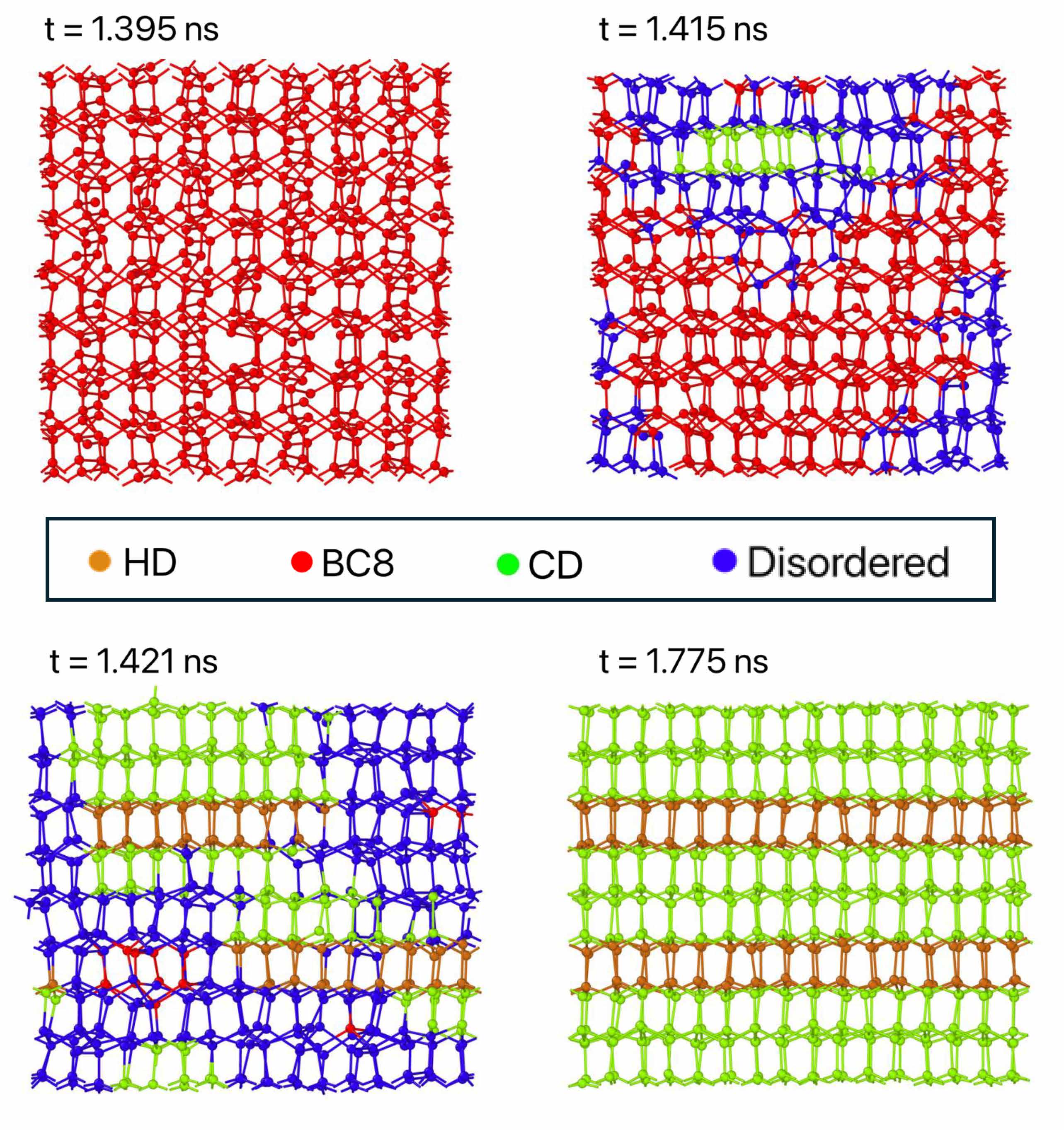}
\caption{Transformation of BC8 to CD and HD upon decompression
to 200 GPa and 3000 K. Initial formation of the CD structure
occurs at 1.415 ns, followed by its growth and creation 
of a final structure consisting
of alternating CD and HD layers after about 8 ps.  The time evolution of the phase fractions is shown in Fig.~S7. }
\label{fig:bc8_to_cdhd}
\end{figure}

Finally, we investigated the key question of BC8 metastability upon quenching to ambient conditions. Early theoretical work \cite{Mailhiot1991} predicted that, at low pressure, BC8 may preferentially transform to HD rather than CD. To test this, we performed metadynamics simulations using a 1024-atom BC8 supercell at 3000 K and pressures of 500, 300, and 200 GPa. In all cases, the system had to overcome a substantial barrier before transforming into a mixture of CD and HD phases. The transformation pathway observed at 200 GPa and 3000 K is shown in Fig.~\ref{fig:bc8_to_cdhd} and Fig.~S7. The pressure dependence of the barrier (Fig.~S10) suggests that it decreases rapidly below 200 GPa. Accordingly, we carried out a standard NPT MD simulation at 100 GPa and 3000 K, where we observed a spontaneous transformation of BC8 into a CD/HD mixture that initiated immediately and completed within ~750 ps. To probe the quenchability of BC8 at room temperature, we further conducted a 5-ns NPT MD simulation at 300 K and zero pressure. In this case, BC8 was found to be dynamically stable, suggesting the possibility of recovering this phase under ambient pressure provided the temperature remains sufficiently low.

To conclude, our metadynamics simulations revealed multiple kinetic
pathways for the synthesis of the post-diamond BC8 phase in carbon
at terapascal pressures (shown on the phase diagram in Fig.~\ref{fig:kinetic_phase_diag}). We found that pressures around 1.8 TPa and temperatures between 3500 and 5000 K are optimal for direct BC8
formation, starting from both CD and HD diamond. In the latter case, the BC8  crystallization from the disordered intermediate droplet appears to proceed faster. In addition, alternative transformation routes between post-diamond phases at these extreme conditions were explored. BC8 is formed upon decompression of the SC phase, and might be quenchable to ambient pressure at room temperature.  From a methodological
standpoint, this study demonstrates that metadynamics employing simple generic
collective variables such as coordination numbers, combined with
quantum-accurate nachine learning potentials is a powerful approach
for uncovering novel phase transformation mechanisms in complex materials.
It enables the resolution of atomistic details of nontrivial nucleation
processes under high-temperature conditions, where entropic contributions
are substantial and cannot be neglected.

\begin{acknowledgments}
R.M. is supported by the Slovak Research and Development Agency under
Contracts No.~APVV-19-0371 and No.~APVV-23-0515. Part of the calculations was performed on the GPU TITAN V provided by the NVIDIA grant. Part of the research
results was also obtained using the computational resources procured in
the national project National competence centre for high performance
computing (project code: 311070AKF2) funded by European Regional Development
Fund, EU Structural Funds Informatization of society, Operational
Program Integrated Infrastructure. R.M. acknowledges help of Matej Badin in the initial phases of this project. 
\end{acknowledgments}

The work at USF is supported by Lawrence Livermore National
Laboratory's Academic Collaborative Team award, DOE/NNSA (Award Nos.
DE-NA-0004089 and DE-NA0004234), DOE/FES (Award No. DE-SC0024640),
NSF (Award No. 2421937), and NASA (Award No. 80NSSC25K7172). The computational work at USF is  performed using leadership-class HPC systems: OLCF Frontier at Oak Ridge National Laboratory (ALCC and INCITE Awards Nos. MAT198 and MAT261), ALCF Aurora at Argonne National Laboratory (ALCC Award “FusAblator”), Perlmutter at National Energy Research Scientific Computing Center (NERSC award FES-ERCAPm3993 and  ERCAP0034777) and TACC Frontera at the University of Texas at Austin (LRAC Award No. DMR21006).


\end{document}